\documentstyle[aps,psfig]{revtex}
\newcommand{\be}{\begin{equation}}
\newcommand{\ee}{\end{equation}}

\newcommand{\psib}{\overline{\psi}}

\newcommand{\dd}{\partial\hspace{-7pt}/}

\newcommand{\ber}{\begin{eqnarray}}
\newcommand{\eer}{\end{eqnarray}}
\newcommand{\hsp}{\hspace*{1pt}}

\begin{document}
\draft
\title {\bf Rotating Twin Stars and Signature of Quark-Hadron Phase Transition}
\vskip 0.2in
\author{Abhijit Bhattacharyya$^a$, Sanjay K. Ghosh$^b$, 
Matthias Hanauske$^c$ and Sibaji Raha$^b$}
\address{$^a$ Department of Physics, Scottish Church College, 1 \& 3, Urquhart 
Square, Kolkata - 700 006, INDIA}
\address{$^b$ Department of Physics, Bose Institute, 93/1, A.P.C. Road,
Kolkata - 700 009, INDIA}
\address{$^c$ Institut f\"{u}r Theoretische Physik, J.W.Goethe-Universitat,
D-60054 Frankfurt am Main, GERMANY}
\maketitle
\vskip 0.4in
\begin{abstract}
The quark hadron phase transition in a rotating compact star has 
been studied. The NLZ-model has been used for the hadronic sector 
and the MIT Bag model has been used for the
quark sector. It has been found that rotating twin star (third family) 
solutions 
are obtained upto $\Omega \approx 4000 s^{-1}$. Stars which are rotating 
faster than this limit do not show any twin star solution. A 
backbending in moment of inertia is also observed in the supermassive 
rest mass sequences. The braking index is found to diverge for a star 
having pure quark core. 
\end{abstract}
\vskip 0.3in
One of the best possible laboratories to study strongly interacting 
matter at super-nuclear densities is the compact stars \cite{a1}. The matter 
density near the core of a compact star can be even $10$ times 
that of normal nuclear matter. At such a high density different exotic phase
transitions may take place in the strongly interacting matter \cite{phr}. 
Some of 
them are the quark-hadron phase transition, the kaon condensation 
and hyperonic phase transition \cite{a2}. There has been a large number of 
efforts to correlate these phase transitions with the observable 
properties of a compact star. However, all the stars, observed so far, 
can be explained within the periphery of strongly interacting nuclear 
matter without invoking any exotic phase transition. 

For a static (non-rotating) compact star the one possible signature is to 
look at the phase transition is the mass-radius relationship. However,
the situation is little bit different for a rotating star. It has been 
argued by Glendenning {\it et. al.} \cite{a3} that one of the 
important quantity 
to probe, in order to study the phase transition, is the braking index. 
The central density of a rapidly rotating pulsar increases with time as 
it spins down and when a certain central density is achieved a quark
phase may appear at the core of the star. Glendenning 
{\it et. al.} \cite{a3} found that when the pure quark core 
appears the star undergoes 
a brief era of spin-up and the braking index shows an anomalous behaviour.
Furthermore, when the moment of inertia was studied as a function of angular 
velocity it showed a "backbending".

Motivated by the above results some authors have studied the backbending 
phenomenon using different EOS and also with different rotating star codes.
Recently Spyrou and Stergioulas \cite{a4}  have published an 
interesting result. They 
have studied the rest mass sequences for a particular EOS (with a 
quark-hadron phase transition) using their code {\bf "rns"}. They argued that 
previous results needed some numerical refinements and concluded that 
backbending is observed only for the supermassive sequences. 

Recently some of us have used the NLZ model (which is a 
variant of the non-linear Walecka model) for the hadronic sector and 
MIT Bag model for the quark sector to look at the quark-hadron phase 
transition in static compact stars \cite{a5}. In that work it was found that 
there was a solution for the third family of stars known as twin 
stars \cite{a6}. 
In this article we employ the same EOS for a rapidly rotating star. 
The basic motivation is to study the fate of the twin stars in a rotating 
model and also to study the possible signatures of these stars. First 
we will describe the model that we use here. Then the General Relativistic 
features of a rotating star will be briefly outlined and at the end we will
discuss the results.

In this paper we will use one variant of the non-linear Walecka 
model, called the NLZ model \cite{a7}, for the hadronic sector and the 
MIT Bag model for the quark sector. 

The lagrangian density for the hadronic model that we use is given
 by \cite{a7}:
\begin{equation}
{\cal L} = {\cal L}_0 + {\cal L}_{YY} + {\cal L}_l
\end{equation}
where

\begin{eqnarray}
{\cal L}_0&=&\sum_B \psib_B\left(i\dd-m_B \right) \psi_B + {1 \over 2} 
\partial^\mu \sigma \partial_\mu \sigma - U \left(\sigma \right) 
-{1 \over 4} G^{\mu \nu} G_{\mu \nu} + U \left(\omega \right) \nonumber\\
&-&{1 \over 4} {\vec B}^{\mu \nu} {\vec B}_{\mu \nu} 
+ 
{1 \over 2} m_\rho^2 {\vec R}^\mu {\vec R}_\mu
- \sum_B \psib_B \left(g_{\sigma B} \sigma + g_{\omega B}\omega^\mu\gamma_\mu 
+ g_\rho{\vec R}^\mu\gamma_\mu {\vec \tau}_B \right)\psi_B
\label{hadlag1}
\end{eqnarray}

\ber
{\cal L}_{YY}&=&{1 \over 2} \left(\partial^\mu \sigma^* \partial_\mu \sigma^* - 
- m^2_{\sigma^*} \sigma^{*2}\right)  - {1 \over 4} S^{\mu \nu} S_{\mu \nu} 
+{1 \over 2} m_\phi^2 \phi^\mu \phi_\mu \nonumber \\
&-& \sum_B \psib_B \left(g_{\sigma^* B} \sigma^* 
+ g_{\phi B} \phi^\mu\gamma_\mu \right)\psi_B
\eer

\be
{\cal L}_l=\sum_{l=e,\mu} \psib \left(i\dd-m_l \right) \psi_l 
\ee

In the above equations the $\sum_B$ runs over all the baryons ($p, n, 
\Lambda, \Sigma^0, \Sigma^+, \Sigma^-, \Xi^0$ and $\Xi^-$) and the 
$\sum_l$ runs over all the leptons (electrons and muons). The piece 
of the Lagrangian ${\cal L}_{YY}$ is responsible for the hyperon-hyperon 
interactions \cite{a7}. The meson fields are 
$\sigma, \omega, {\vec R}, \sigma^* 
(f_0(975))$, and $\phi$. The sigma and omega meson potentials are given 
by \cite{a7,a9,a10}

\ber
U_\sigma&=&\frac{(m_\sigma\sigma)^2}{2}
+\frac{g_2\hsp\sigma^3}{3}+\frac{g_3\hsp\sigma^4}{4}\,,\label{spot}\\
U_\omega&=&\frac{m_\omega^2 \hsp\omega^\mu \omega_\mu}{2}+
\frac{g_{3\omega}\hsp(\omega^\mu \omega_\mu)^2}{4}\,.\label{vpot}
\eer

The nucleon coupling 
constants are chosen from the fit to finite nuclei properties. The 
vector coupling constants of the hyperons are chosen according to the 
SU(6) symmetry as \cite{a11,a12,a13,a14}
\ber
{1 \over 3} g_{\omega N} = {1 \over 2} g_{\omega \Lambda} = 
{1 \over 2} g_{\omega \Sigma} = g_{\omega \Xi} \nonumber\\
g_{\rho N} = {1 \over 2} g_{\rho \Sigma} = g_{\rho \Xi} \nonumber\\
g_{\rho \Lambda} = 0 \nonumber\\
2 g_{\phi \Lambda} = 2 g_{\phi \Sigma} = g_{\phi \Xi} 
= - {{2 {\sqrt {2}}} \over 3} g_{\omega N} \nonumber \\
g_{\phi N} = 0 
\eer
The hyperonic scalar coupling constants are chosen to reproduce the measured 
values of the optical potentials \cite{a15,a16}, 
\ber
U^{\left(N\right)}_\Lambda = U^{\left(N\right)}_\Sigma =  -30 MeV, 
\hspace*{10pt} U^{\left(N\right)}_\Xi = -28 MeV \nonumber \\
U^{\left(\Xi\right)}_\Xi = U^{\left(\Xi\right)}_\Lambda = 
2 U^{\left(\Lambda\right)}_\Xi = 2 U^{\left(\Lambda\right)}_\Lambda = -40 MeV
\eer

We use the parameter sets as given in ref.\cite{a5}

At the mean-field level, the meson fields are replaced by their ground 
state expectation values. The equations of motion for different meson fields 
then can be  obtained by standard methods,
they are  given by 
\ber
m_\sigma^2 \sigma + {{\partial} \over {\partial \sigma}} U(\sigma) 
= \sum_B g_{\sigma B} \rho_s^B \nonumber \\
m_{\sigma^*}^2 \sigma^*  = \sum_B g_{\sigma^* B} \rho_s^B \nonumber \\
m_\omega^2 \omega_0 + g_{3 \omega} \omega_0^3
= \sum_B g_{\omega B} \rho_V^B \nonumber \\
m_\rho^2 R_{3,0} = \sum_B g_{\rho B} \tau_3^B \rho_V^B \nonumber \\
m_\phi^2 \phi_0 = \sum_B g_{\phi B} \tau_3^B \rho_V^B 
\eer
The single particle energies for baryons follow from the Dirac equation 
\be
E_B(k) = g_{B\omega}\omega_0 + g_{B\phi}\phi_0 +
 g_{B\rho}\tau_3^B\rho_0 + \sqrt{k^2+m_B^{*2}}
\ee
where the baryon effective masses are given by 
\be
m_B^* = m_B + g_{\sigma B} \sigma + g_{\sigma^* B} \sigma^* 
\ee
For leptons the energy is $E(k) = \sqrt{k^2+m_l^{*2}}$ and we work with 
the vacuum masses of the leptons. 

The pressure and energy density obtained from these models can be written 
as \cite{a5}
\ber
\epsilon &=& \frac{1}{2}m_\sigma^2 \sigma^2
+ \frac{g_2}{3}\sigma^3 + \frac{g_3}{4}\sigma^4
+ \frac{1}{2}m_{\sigma^*}^2 {\sigma^*}^2
+ \frac{1}{2}m_\omega^2 \omega_0^2 + \frac{3}{4} g_{3\omega} V_0^4
\cr && {}
+ \frac{1}{2}m_\rho^2 R_{0,0}^2
+ \frac{1}{2}m_\phi^2 \phi_0^2
+ \sum_{i=B,l} \frac{\nu_i}{(2\pi^3)} \int_0^{k_F^i} d^3 k
\sqrt{k^2 + {m^*_i}^2} \\ \cr
P &=& - \frac{1}{2}m_\sigma^2 \sigma^2
- \frac{g_2}{3}\sigma^3 - \frac{g_3}{4}\sigma^4
- \frac{1}{2}m_{\sigma^*}^2 {\sigma^*}^2
+ \frac{1}{2}m_\omega^2 \omega_0^2 + \frac{1}{4} g_{3\omega} V_0^4
\cr && {}
+ \frac{1}{2}m_\rho^2 R_{0,0}^2
+ \frac{1}{2}m_\phi^2 \phi_0^2
+ \sum_{i=B,l} \frac{\nu_i}{3(2\pi^3)} \int_0^{k_F^i} d^3 k
\frac{k^2}{\sqrt{k^2 + {m^*_i}^2}}
\eer
where $\nu_i$ is the degeneracy factor of the $i$-th species.
This EOS will be used to look at the star properties. 

For the quark sector we use the MIT Bag model \cite{bag}.  
The pressure and energy density of the quark matter is given 
by 
\be
\epsilon^Q = \sum_{f=u,d,s} {{\nu_f} \over {2 \pi^2}} 
\int_0^{k_F^f} dk k^2 {\sqrt {m_f^2 + k^2}} + B
\ee

\be
P^Q = \sum_{f=u,d,s} {{\nu_f} \over {6 \pi^2}} 
\int_0^{k_F^f} dk k^4 {{k^4} \over {\sqrt {m_f^2 + k^2}}} - B
\ee

In the above two equations the sum runs over the three flavours of quark 
and $\nu_f$ is the degeneracy.

In this work the deconfinement phase transition is assumed to be of 
first order which proceeds via mixed phase. At zero temperature the 
mixed phase should follow Gibbs criterion \cite{a3} in presence of two 
conserved charges. The Gibbs criterion dictates 

\ber 
P^H(\mu_b,\mu_e) = P^Q (\mu_b,\mu_e) \nonumber\\
\mu_b = \mu_b^H = \mu_b^Q \\
\mu_e = \mu_e^H = \mu_e^Q
\eer
where $P^H$, $\mu_b^H$ and $\mu_e^H$ are the hadronic pressure, hadronic 
contribution to the baryon chemical potential and hadronic contribution to 
the charge chemical potential respectively. The similar quantities for the 
quark phase are denoted as $P^Q$, $\mu_b^Q$ and $\mu_e^Q$ for the quark 
phase. 

The volume averaged energy density in the MP can be written as 
\be
\epsilon = (1-\lambda) \epsilon^H (\mu_b,\mu_e) + \lambda 
\epsilon^Q (\mu_b,\mu_e) 
\ee
where $\lambda$ is fraction of quark matter present in the mixed phase. 

In the quark sector, we have taken light quark masses to be zero and 
strange quark mass to be 150 MeV. So bag pressure \( B \) is the only parameter. The 
hadron-quark phase 
transition is possible for \( B^{1/4} \) in the range 175 - 190 MeV. 
But twin star solution 
is obtained for a narrow range for \( B^{1/4} \approx \) 180 - 182 MeV. In the present 
letter we have given the results for \( B^{1/4} \) = 180 MeV.
    
The EOS is plotted in figure 1. From the figure one can see that 
the phase transition starts at around $\epsilon_c = 4 \times 10^{14} gm/cm^3$ 
and it 
ends around $\epsilon_c = 1 \times 10^{15} gm/cm^3$. We would like to 
mention at 
this stage that due to the phase transition there is a substantial change 
in the slope (sound speed) of the eos. This point will be important 
in the context of  our results discussed later.

Once the EOS is obtained the next job is to solve the Einstein's 
equations for the rotating stars using the EOS. To solve the Einstein's 
equations we follow the procedure adopted by Komatsu {\it et.al.} \cite{a17}.
 In this 
work we briefly outline some of the steps only. The metric for a 
stationarily rotating star can be written as \cite{a18}
\be
ds^2 = -e^{\gamma+\rho} dt^2 + e^{2\alpha} \left(dr^2 + r^2 d\theta^2\right)
+e^{\gamma-\rho} r^2 sin^2\theta \left(d\phi - \omega dt\right)^2
\ee
where $\alpha$, $\gamma$, $\rho$ and $\omega$ are the gravitational potentials
which depend on $r$ and $\theta$ only. The Einstein's equations for 
the three potentials
$\gamma$, $\rho$ and $\omega$ have been solved by Komatsu {\it et.al.} using 
Green's function technique. The fourth potential $\alpha$ has been determined 
from other potentials. All the physical quantities may then be determined 
from these potentials \cite{a18}. 

Solution of the potentials, and hence the calculation of physical quantities, 
is numerically quite an involved process. There are several numerical codes 
in the community for this purpose. In this work we use the {\bf 'rns'} code. 
This code developed by Stergioulas is a very efficient   
in calculating the rotating star properties. We do not mention the details
of the code here; they may be obtained in ref. \cite{a17,a18}. 
We have used this code to study the rest mass sequences 
and the $\Omega$ sequences for both normal and supermassive stars. 

In figure 2 we show the mass-radius relationship for 
different values of $\Omega$ starting from the static limit to $\Omega = 
5500 s^{-1}$. From the plots one can see that as the value 
of $\Omega$ increases 
both the radii and the masses increase, which is an obvious result. However  
we find some interesting results. The static mass radius plot 
shows that there are occurrences of twin stars in this model. 
The same interesting 
result, {\it i.e.} the occurrence of twin stars, is also observed for rotating 
stars. However, as we increase $\Omega$, the third family solution becomes 
less probable as can be seen from the plots. The third family solutions 
are obtained until the rotational velocity is $\approx 4000 s^{-1}$. For 
stars rotating at higher angular velocity, the third family of stars cannot 
be observed. 

In figure 3 we have plotted the moment of inertia ($I$) as a function of 
$\Omega$ for different rest masses {\it i.e.} for both normal and supermassive 
sequences. For the normal sequence the moment of inertia increases with the 
angular velocity monotonically. However it is not exactly the same for the 
supermassive sequences. For a supermassive sequence, there is one branch 
of the curve for which the moment of inertia increases with angular velocity 
but there is another branch for which the moment of inertia decreases with 
increase in $\Omega$. This phenomenon is known as backbending. As pointed out 
by several authors, this phenomenon could be a possible signature of the 
quark-hadron phase transition. Here we would like to emphasize that even for 
supermassive sequence, the backbending is only observed for the cases where 
density in the core is high enough to facilitate pure quark phase. For example 
though the range of rest mass $1.35 M_\odot < M_0 < 1.48 M_\odot$ 
lies in the supermassive domain it does not have a pure quark core and 
it also does not show any back bending. 
 
In the next figure {\it i.e.} in figure 4 the angular momentum is 
plotted as a function of energy density for different rest masses. We have 
plotted these curves only for the supermassive sequences. From the 
plots we see that the angular momentum ($J$) decreases with $\epsilon_c$ 
then it increases (which is the unstable region) and then again decreases 
showing a third family solution. 

In figure 5 the braking index has been plotted against $\Omega$. As 
pointed out by Glendennig {\it et al} the divergence of braking index is 
a signature of the phase transition. This quantity is defined as
\be
\nu (\Omega) = 3 - {{3I^{'}\Omega + I^{''}\Omega^2} \over {2I + I^{'}\Omega}}
\ee

The 
braking index is found to 
diverge for supermassive sequences whereas no fluctuation has been 
observed for the normal sequence. The amount of the fluctuations 
is a manifestation of the size of the quark core of the star. For a star 
in the normal sequence, the central density is such that the core of the 
star lies at the mixed phase and hence the braking index does not show any 
fluctuation. For the supermassive sequence, in the stars for which the 
central density is high enough to accommodate a quark core, the 
braking index diverges. For the normal sequence that we have plotted here 
the maximum central density is $\sim 8 \times 10^{14} gm/cm^3$ whereas 
for the supermassive sequence it is $\sim 2 \times 10^{15} gm/cm^3$. 
Comparing these numbers with the different central energy densities as 
shown in figure 1 it is quite obvious that for the normal sequence the core 
is in the mixed phase region whereas for the supermassive sequence the core 
of the star consists of pure quark matter.

To summarise, we have studied the occurence of twin solutions in rotating 
compact stars. The third family solution or twin stars occur due to a 
substantial change in the sound speed. We have found that though the twin 
solutions are present in rotating stars they vanish after a certain 
value of $\Omega$. Hence stars which are rotating with very high angular
velocity do not show a twin solution. There is another interesting observation 
in this work compared to that of Spyrou {\it et.al.} \cite{a4}. They have 
found that 
there is a small oscillation of the braking index in the normal sequence. 
However we do not find any such oscillation in the normal sequence. This 
may be due to the fact that in the central region we have only a mixed 
phase whereas they have a quark core in the central region. This is obviously 
ascribable to the difference in the EOS. 

{\bf Acknowledgements}
AB would like to thank University Grants Commission for partial support 
through the grant PSW-083/03-04.

\newpage
\begin{figure}[t]
\vskip-2.1cm
\hskip-1.76cm
\centerline{\psfig{file=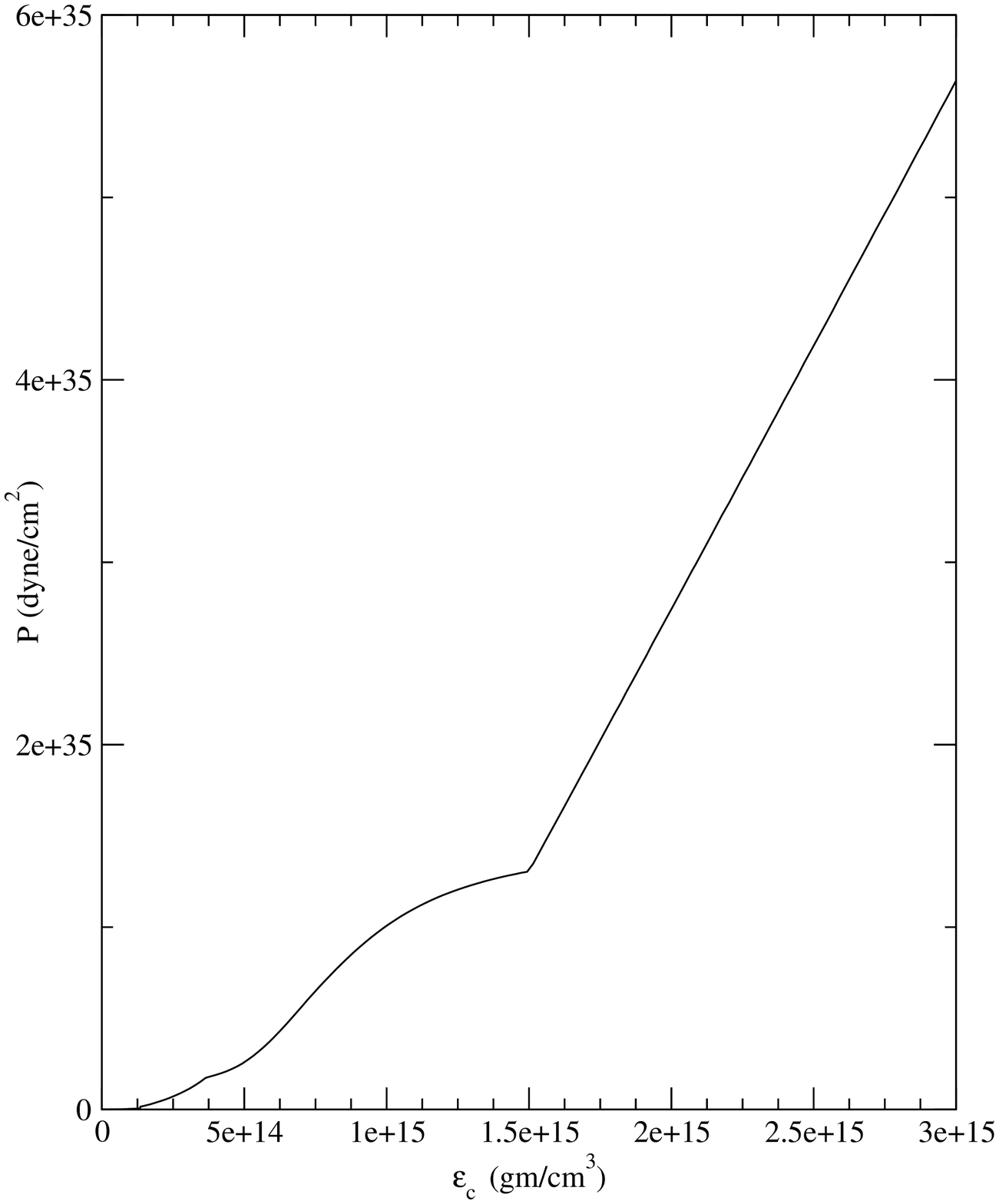,width=15cm}}
\caption{Equation of state of the model considered here.}
\end{figure}
\newpage
\begin{figure}[t]
\vskip-2.1cm
\hskip-1.76cm
\centerline{\psfig{file=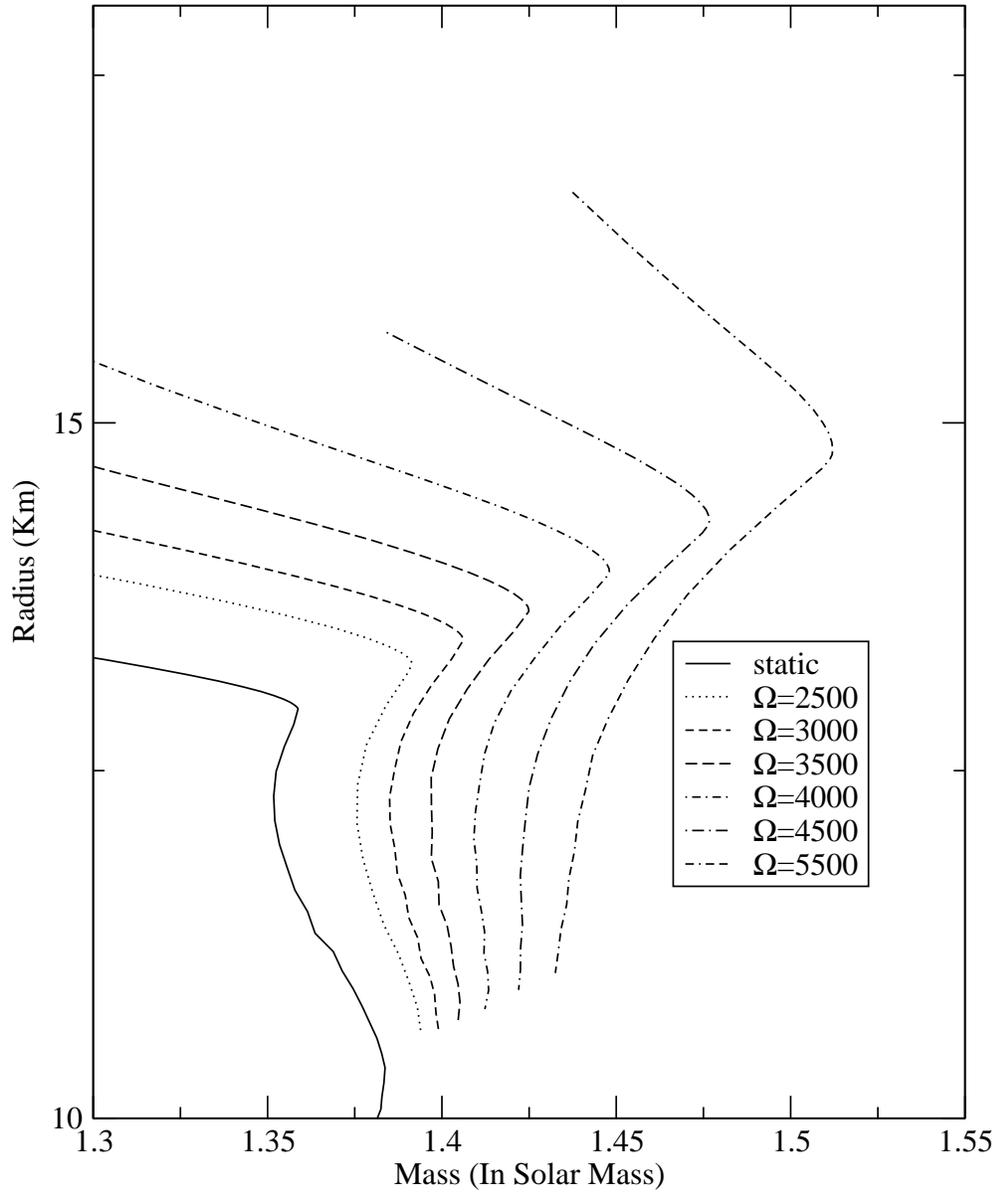,width=15cm}}
\caption{Radius-Mass plots for different $\Omega$.}
\end{figure}
\newpage
\begin{figure}[t]
\vskip-2.1cm
\hskip-1.76cm
\centerline{\psfig{file=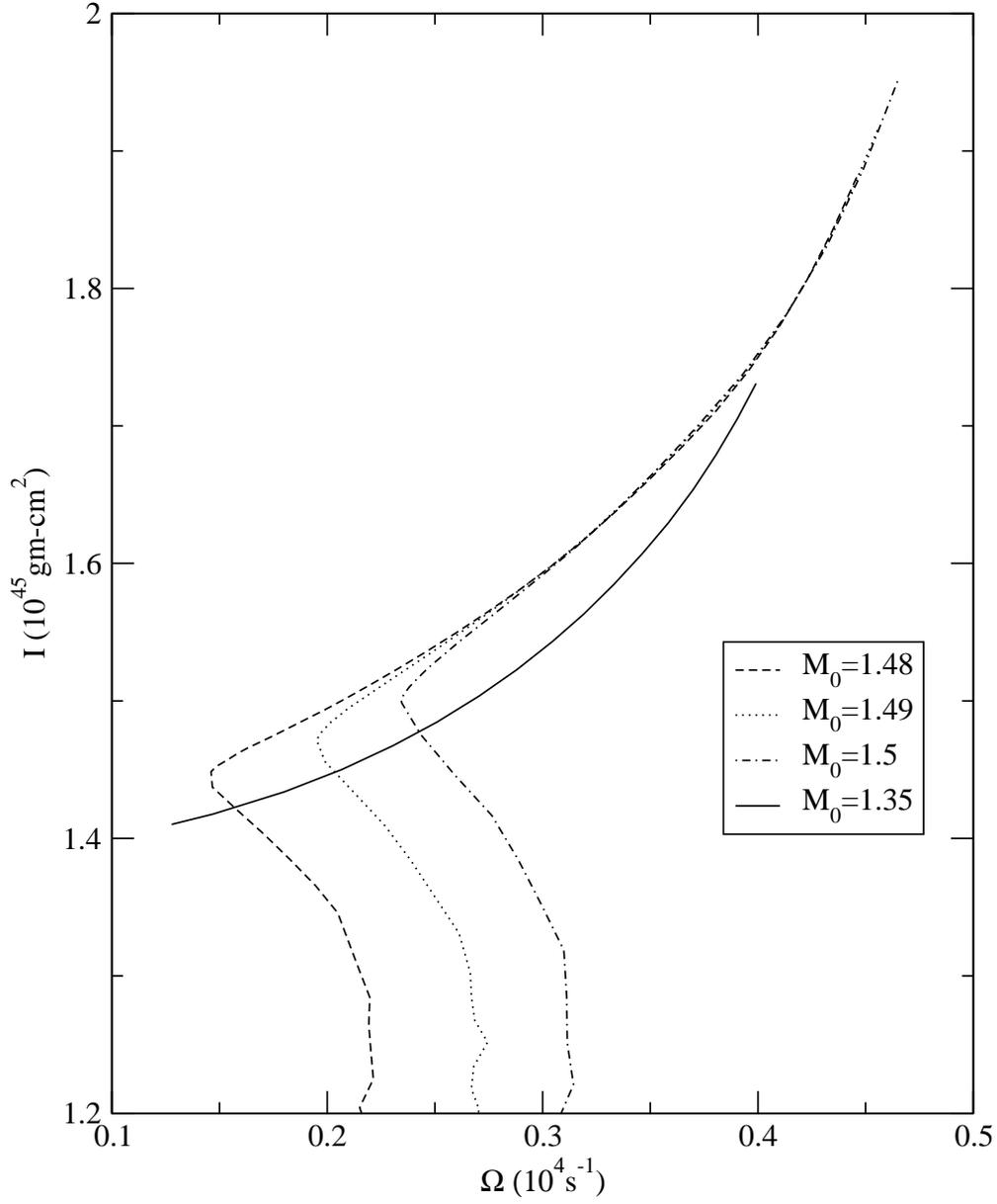,width=15cm}}
\caption{Moment of inertia as a function of $\Omega$ for different rest mass.}
\end{figure}
\newpage
\begin{figure}[t]
\vskip-2.1cm
\hskip-1.76cm
\centerline{\psfig{file=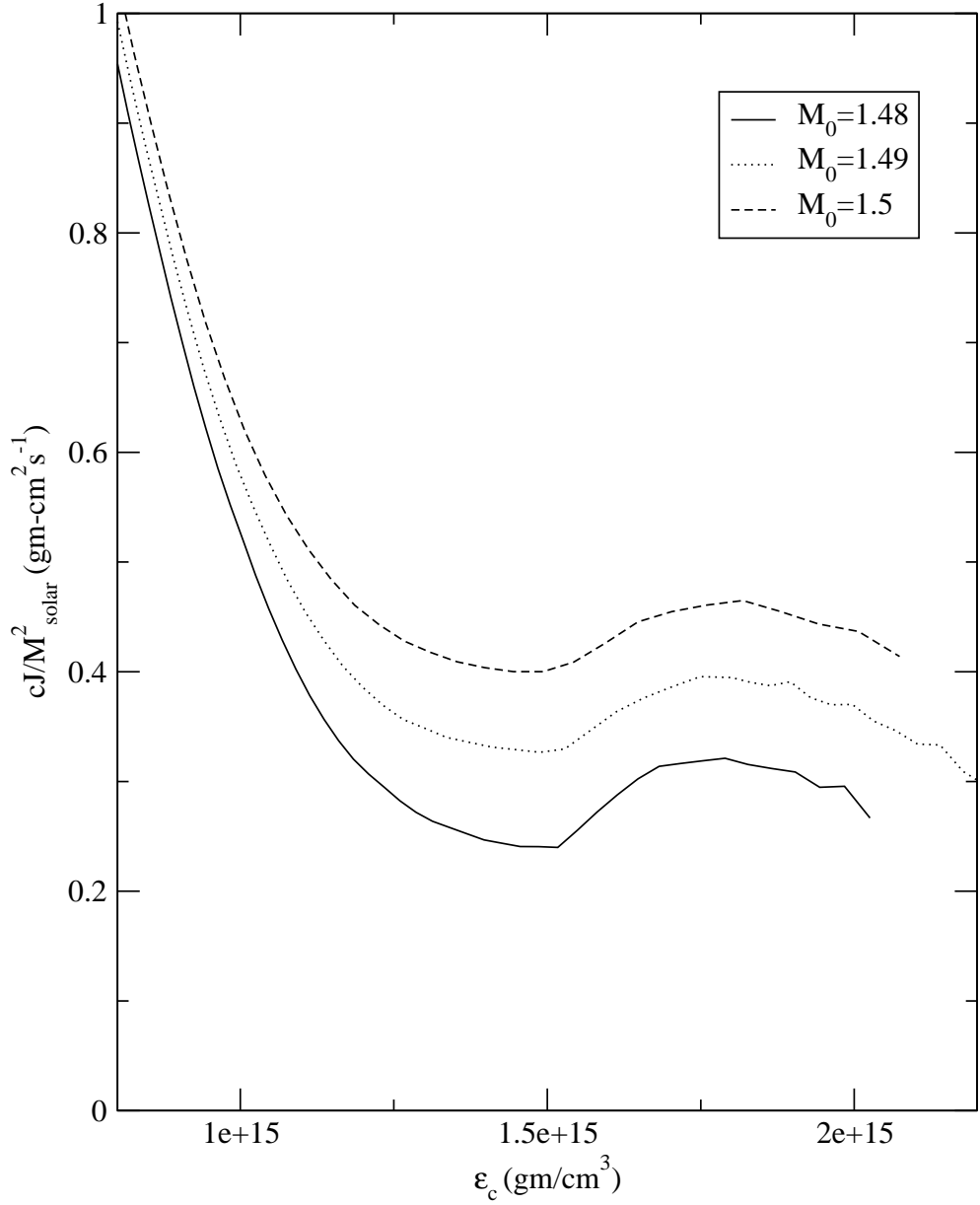,width=15cm}}
\caption{J as a function of $\epsilon_c$ for different rest mass.}
\end{figure}
\newpage 
\begin{figure}[t]
\vskip-2.1cm
\hskip-1.76cm
\centerline{\psfig{file=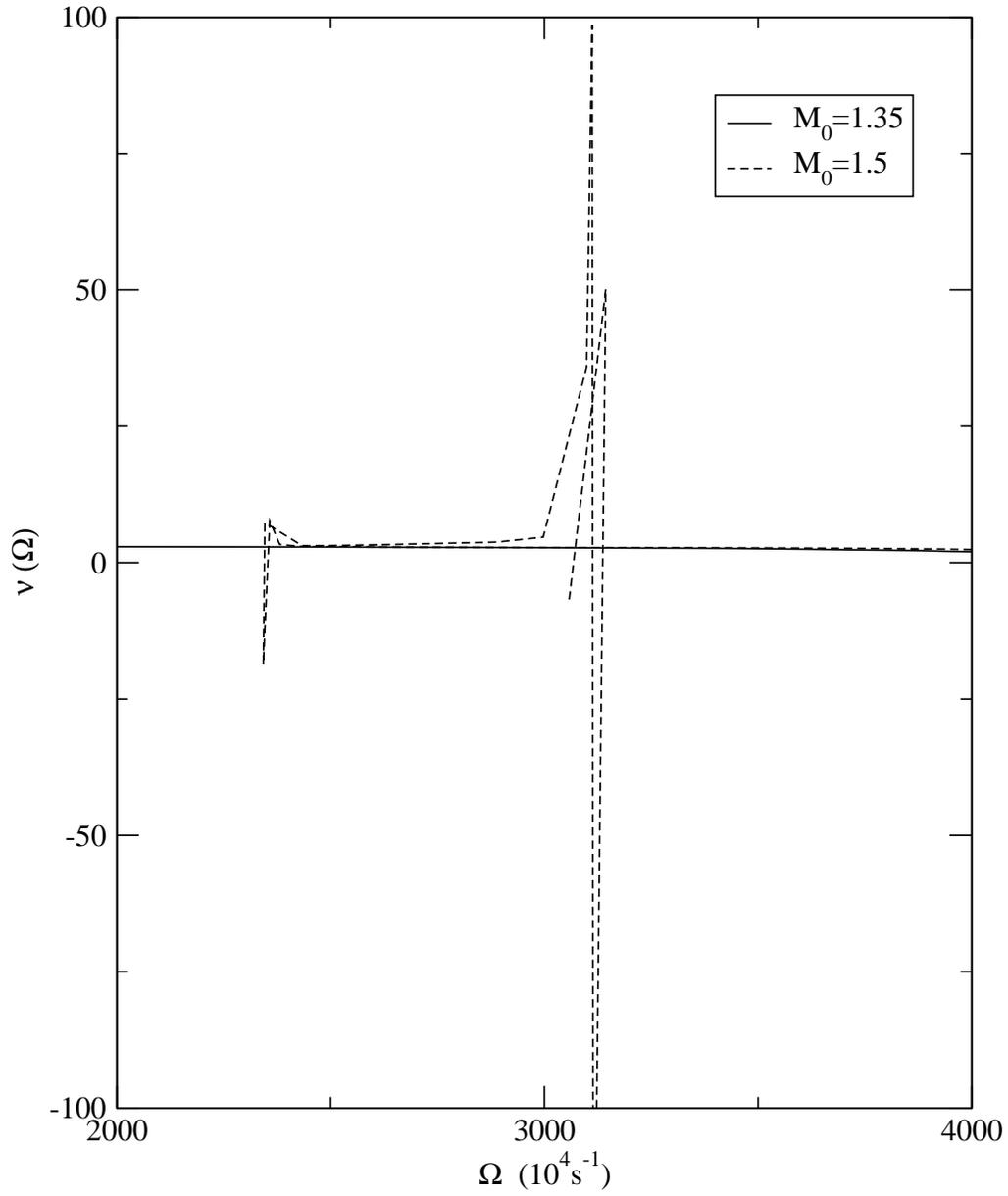,width=15cm}}
\caption{Braking index as a function of $\Omega$.}
\end{figure}
\end{document}